\documentclass[10pt]{article}
\usepackage{graphicx}
\usepackage{cite,./mcite}
\usepackage{amsmath,amssymb}

\usepackage{fancyhdr}

\begin{document}
\pagestyle{fancyplain}
\rhead{\begin{large}Edinburgh 2007-19 \end{large}}
\title{\textbf{BFKL at NNLO}}
\author{S.~Marzani $^1$\protect\footnote{~~~Talk presented at $12^{th}$ International Conference on Elastic and Diffractive Scattering: 
\emph{Forward Physics and QCD}, DESY, Hamburg, Germany, 21-25 May 2007.} , R.D.~Ball $^1$, P.~Falgari $^2$, S.~Forte $^3$ 
\vspace{0.8cm} \\
$^1$ \emph{School of Physics, University of Edinburgh,} \\
\emph{  Edinburgh EH9 3JZ, Scotland} \vspace{0.5cm}\\
$^2$ \emph{Institut f\"ur Theoretische Physik E, RWTH Aachen,} \\
\emph{ D-52056 Aachen, Germany} \vspace{0.5cm} \\
$^3$ \emph{Dipartimento di  Fisica, Universit\`a di Milano and} \\
\emph{INFN, Sezione di Milano,} \\ \emph{Via Celoria 16, I-20133 Milan, Italy}}
\date{}
\maketitle
 \begin{abstract}
\begin{normalsize}We present a recent determination of an approximate expression for
the $O(\alpha_s^3)$ contribution $\chi_2$ to the kernel of the
BFKL equation. This includes all collinear and
anticollinear singular contributions and is derived using
duality relations between the GLAP and BFKL kernels.
\end{normalsize}
\end{abstract}
\newpage
\pagestyle{plain}
\rhead{}
\section{QCD at small-$x$}
Reducing the theoretical uncertainties in cross-sections for hadron
colliders
requires the computation of higher order contributions in
perturbative QCD, both at fixed order and at the resummed level.
In particular at high energy colliders such as the LHC, 
we must be able to control both the logarithms of $Q^2$ and $x$ as given by
GLAP and BFKL evolution. Fixed--order BFKL kernels, which
resum only logs of $x$, have been widely used  in many studies (such as
saturation and BFKL Monte Carlos for final states). The BFKL kernel
has been computed explicitly at the next-to-leading order accuracy
\cite{chi1}; here we present an approximation of the NNLO contribution.
The fixed--order expansion of the kernel is known to be slowly
convergent, hence the NNLO contribution is important for an
accurate assessment of the NLO uncertainty at any particular scale.

Let us consider the GLAP and the BFKL equations:
   \begin{eqnarray}\label{glap}
         \frac{d}{d t}G(N,t)&=& \gamma(N,\alpha_s)G(N,t)\,,
         \\
\label{bfkl}
         \frac{d}{d \xi}G(\xi,M)&=& \chi(M,\alpha_s)G(\xi,M)\,,
 \end{eqnarray}
 They describe, respectively, the evolution with respect to 
 $t=ln \frac{Q^2}{\mu^2}$
 and $\xi = ln \frac{s}{Q^2} =ln\frac{1}{x}$ 
 of the singlet parton density. The complex variables $N$ and $M$ are the
Mellin moments with respect to $x$ and $Q^2$ respectively: upon taking moments the integro-differential evolution equations
become ordinary differential equations. Note that  
 the GLAP evolved parton density $G$ is integrated
 over the transverse momenta, while the BFKL equation is 
 usually written in terms of the unintegrated
 quantity $\mathcal{G}$. We shall return to this issue in the next
 section.

 Eq. (\ref{bfkl}) is written in the fixed coupling approximation; the introduction of the running of the coupling is nontrivial
because upon Mellin transform $\alpha_s(t)$ becomes a differential
operator:
\begin{eqnarray*}
   \alpha_s(t)=\frac{\alpha_s}{1+\alpha_s\beta_0 t}
   &\Longrightarrow&
   \hat{\alpha}_s=\frac{\alpha_s}{1-\alpha_s\beta_0 \frac{\partial}{\partial M}} \,.
\end{eqnarray*}
As a consequence 
the BFKL kernel $\chi(\hat{\alpha}_s,M)$ becomes an operator
beyond leading order. It is useful to notice that different
arguments for the running coupling correspond to different
orderings of the operators.

The fixed-order expansion of the BFKL kernel is:
\begin{equation}
\chi(M,\hat{\alpha}_s)=\hat{\alpha}_s \chi_0(M)+
\hat{\alpha}^2_s\chi_1(M)+ \hat{\alpha}^3_s\chi_2(M)+\dots
\end{equation}
It is well known that the NLO order term $\chi_1$ is large and has
a qualitatively different shape to the leading order kernel $\chi_0$. 
The determination of the NNLO contribution $\chi_2$ is motivated not only by
the slow convergence of the perturbative expansion
but also by the  expectation that the NNLO approximation will have
the same qualitative shape as the LO thus having better stability
proprieties than the NLO.  We compute an approximation to the NNLO
kernel which includes all the collinear $(M \sim 0)$ and
anticollinear $(M \sim 1)$ singularities: this computation is
based on so called duality relations between the BFKL kernel and
the GLAP anomalous dimension.
\section{The collinear approximation of the BFKL \newline kernel}
At fixed coupling, the two
evolution equations (\ref{glap}) and (\ref{bfkl}) admit the 
same leading twist solution when the kernels
are related by:
\begin{equation} \label{dual}
       \chi(\gamma(N,\alpha_s),\alpha_s)= N,  \qquad\qquad
        \gamma(\chi(M,\alpha_s),\alpha_s)= M,
          \end{equation}
and the boundary conditions are suitably matched \cite{1997},
\cite{2000}.

The GLAP anomalous dimension $\gamma(N,\alpha_s)$ has been computed up to
NNLO, i.e. $O\left(\alpha_s^3\right)$ \cite{vogt}. Using duality 
it is
thus possible to determine the first three coefficients of the Laurent
expansion about $M=0$ of the BFKL kernel. This means that we can 
compute {\em all}
the collinear singularities of the $O\left(\alpha_s^3\right)$
contribution: writing
\begin{equation} \label{laurent}
 \chi_2(M)=
\frac{c_{2,-3}}{M^{3}}+\frac{c_{2,-2}}{M^{2}}+\frac{c_{2,-1}}{M}+\ldots
\,.
\end{equation}
the determination of the coefficients $c_{2,-3}$, $c_{2,-2}$ and
$c_{2,-1}$ requires the knowledge of the LO, NLO and NNLO
anomalous dimension respectively.

At LO the calculation is straightforward because it only involves
the inversion of eq. (\ref{dual}), but beyond that several other
contributions must be taken into account.  More precisely  
we have to address the following complications:
\begin{itemize} \item The inclusion of running coupling effects.
\item The relation between kernels for the integrated
and the unintegrated parton density. \item The dependence of the
kernel on the factorisation scheme. \item The choice of kinematic
variables.
\end{itemize}
All these issues were well understood at NLO \cite{salam}, but only
recently under control at NNLO.

The frozen coupling hypothesis is no longer valid beyond leading
order: duality relations still hold but they receive running
coupling contributions \cite{1999}, \cite{2002}. Running coupling
duality has been proved to all orders using an operator method. As
we already noticed the
 running coupling in $M$-space is a differential operator; duality states
that  the BFKL and GLAP solutions coincide if the respective
operator kernels are the inverse of each other when acting on physical
states. Because of non-vanishing commutation relations the
inversions of the kernels is not trivial; the operator formalism
enables us to compute the running coupling corrections in an
algebraic way, calculating commutators of the relevant operators,
e.g.
\begin{equation}
[\hat{\alpha}_s^{-1},M]=-\beta_0+ \alpha_s \beta_0 \beta_1+ \dots
\end{equation} and express the result in terms of
the fixed coupling duals as described extensively in \cite{2006}.

The BFKL equation describes the evolution of a parton density
$\mathcal{G}$ unintegrated over the transverse momenta, while GLAP
of the integrated one $G$. The relation between $\mathcal{G}$ and
$G$ is given by:
 \begin{displaymath}
 \mathcal{G}(N,t)=\frac{d}{dt} G(N,t)\,.
 \end{displaymath}
This gives the following NNLO relation between unintegrated kernel
and the integrated one $\chi^i$ derived from duality:
\begin{equation}
 \chi_2=\chi^i_2-\beta_0\beta_1\frac{\chi^i_0}{M}
-2\beta_0\frac{\chi^i_1}{M}\,.
\end{equation}

The direct computation of the BFKL kernel is based on
determination of the gluon Green's function in the high energy
limit
 in the framework of the $k_{\perp}$- factorisation theorem. This is 
compatible with the usual factorisation theorem of collinear singularities but 
it differs from it by a computable scheme change. This arises from a
 difference in normalisation between of the gluon Green's functions 
 which enter the BFKL equation and GLAP equations. The usual 
computation of the BFKL kernel using gluon Reggeization
\cite{chi1} is performed in the so called  $Q_0$ scheme
\cite{ciaf95}.
 The gluon normalisation factor relating conventional $\overline{MS}$ 
to the $Q_0$ scheme can be factorised as \cite{ciaf05}
 \begin{equation}
 R(N,t) = \mathcal{N}(N,t) \mathcal{R}(N,t)\,,
\end{equation}
where $\mathcal{N}$ contains readily calculable running coupling and
integrated/unintegrated contributions, while $\mathcal{R}$ is
related to the $\overline{MS}$ definition of the anomalous
dimension. The leading log-$x$ contribution to the $\mathcal{R}$
scheme change  was computed in \cite{catani}. We discuss the
collinear approximation of the NLL$x$ scheme change in
\cite{chi2}, where we show that it can be derived from the
analytic continuation of the GLAP anomalous dimension 
to  $d=4- 2 \varepsilon$
space-time dimensions. However although the 
$O(\alpha_s\varepsilon)$ and $O(\alpha_s\varepsilon^2)$ contributions are known 
the $O(\alpha_s^2\varepsilon)$
contribution is not. We assess the uncertainty in our calculation 
due to this unknown contribution to the scheme change in fig.~2 below.

Once we have all the possible contributions which correct duality
relations at NNLO under control, we can compute the collinear
approximation of the BFKL kernel in the $Q_0$ scheme. In such
scheme the result can be extend in the anticollinear region $M\sim
1$ because the kernel is symmetric upon the exchange:
\begin{equation}
 M \leftrightarrow 1-M
\end{equation}
as a consequence of the symmetry of the
 diagrams for BFKL processes upon the exchange of 
the virtualities at the top and the bottom \cite{chi1}, \cite{ciaf99}.
 Before we can exploit this symmetry we must make sure that all
sources of symmetry breaking have been removed.
 The symmetry may be broken by the  choice of kinematic variables 
(e.g. in  DIS we choose $x=\frac{Q^2}{s}$),
 and by the  argument of the running coupling ($\alpha_s(Q^2)$ in DIS). 
The BFKL kernel can be written in symmetric variables
($x_{sym}={\sqrt{Q^2 k^2}}/{s}$) thanks to the relation:
 \begin{equation}
 \chi^{sym}(\hat{\alpha}_s,M)=\chi^{DIS}(\hat{\alpha}_s,M+\frac{1}{2}\chi^{sym}(\hat{\alpha}_s,M))\,.
\end{equation} The kernel in symmetric variables can be extended to the anticollinear
region:
 \begin{equation}
\chi_2^{sym}(\hat{\alpha}_s,M)= \sum_{j=1,5} c_{2,-j} \left[
\hat{\alpha}_s^3\frac{1}{M^j}+\frac{1}{(1-M)^j}\hat{\alpha}_s^3
\right] + O(M^0)\,.
\end{equation}
The different order of the operators in the collinear and
anticollinear regions corresponds to a symmetric choice for the
running coupling. After the symmetrisation one can express the
results canonically ordered with all the powers of
$\hat{\alpha}_s$ on the left. This choice will, of course, break
the symmetry of the kernel.

\begin{figure} \label{plot1}
\begin{center}
\includegraphics[width=10 cm,height=7 cm]{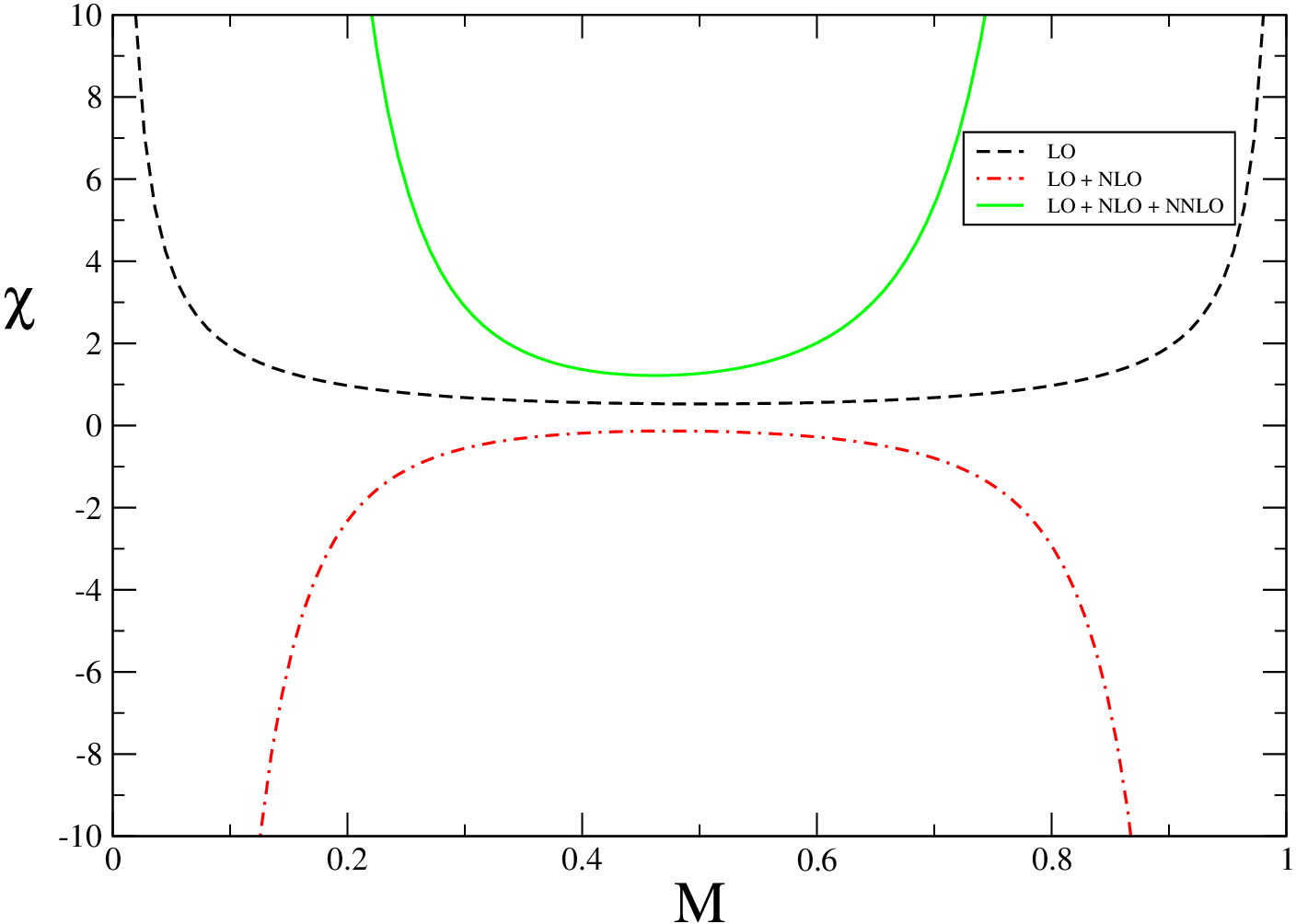}
\caption{This plot shows the LO, NLO and NNLO approximations to the BFKL
kernel in $Q_0$-scheme for $\alpha_s=0.2$.}
\end{center}
\vskip-0.5truecm
\end{figure}

 The results are plotted in
figure $1$. It is clear that the expansion of BFKL kernel is not
well behaved (due to the collinear and anticollinear poles 
at $M=0$ and $M=1$ of increasing order and alternating
sign). However, as expected because of the sign of the dominant
pole, the BFKL kernel at NNLO has a minimum for every value of the
coupling. In figure $2$ we plot the intercept, defined as the
value of the kernel in its minimum, as a function of the coupling
constant.  The inclusion of the NNLO contribution improves the
convergence of the perturbative expansion, however for values of
the coupling constants relevant for phenomenology  (say $\alpha_s
\gtrsim  0.1$) the series has yet to converge.

\begin{figure} \label{plot2}
\begin{center}
\includegraphics[width=10 cm,height=7 cm]{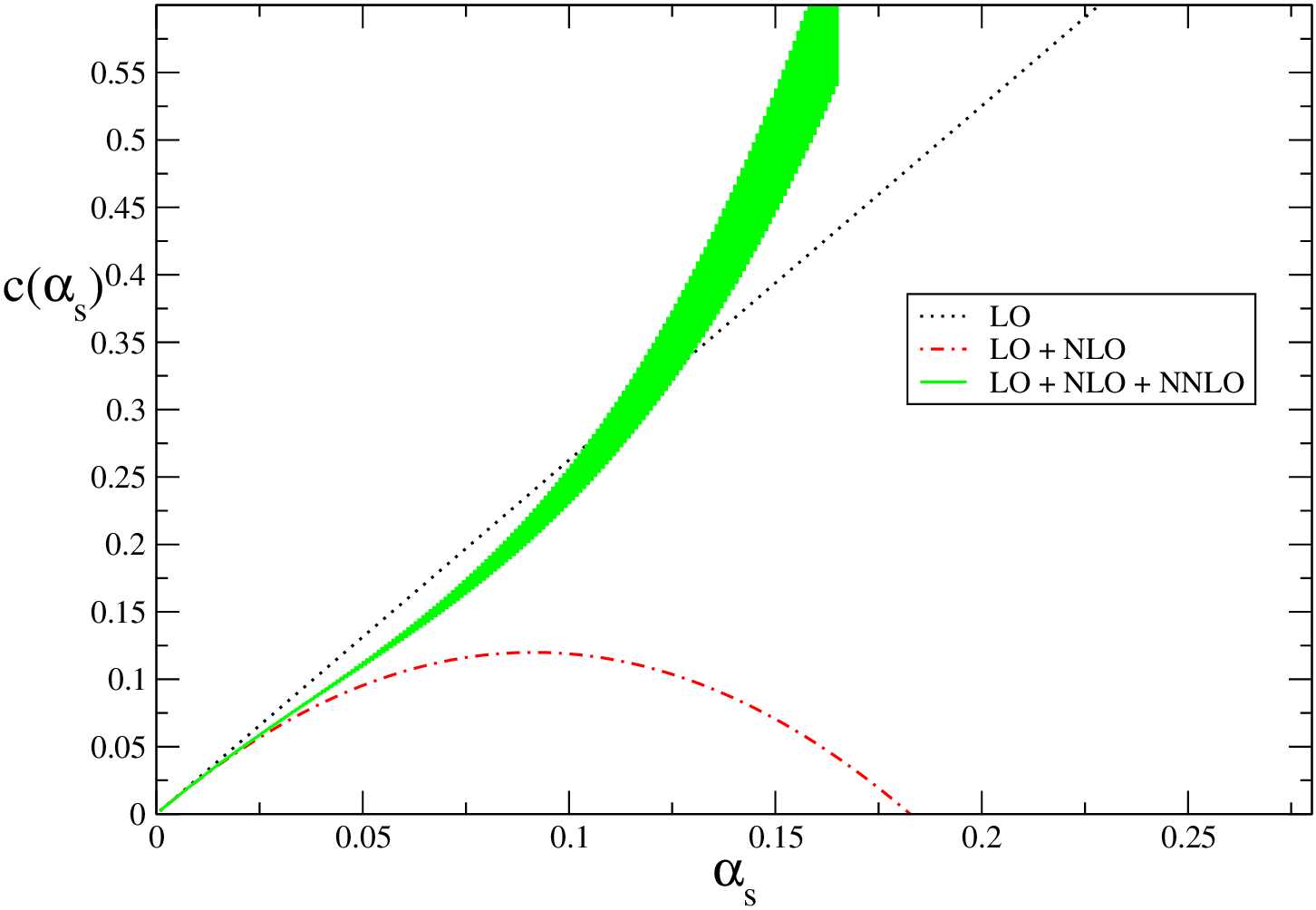}
\caption{This plot shows the intercept as a function of the
coupling, at LO, NLO and NNLO. The error band is due to the 
unknown term in the scheme change.}
\end{center}
\vskip-0.5truecm
\end{figure}

\section{Discussion}

We have seen that thanks to duality relations and the
computation of the anomalous dimension at NNLO, the calculation of
the collinear approximation of the BFKL kernel at $O(\alpha_s^3)$
can be performed.  Here we discuss the accuracy and
the limitations of our result.

We have computed an approximation to the forward BFKL kernel, which has
been azimuthal averaged over the transverse momenta.  The
collinear approximation is based on the computation of
coefficients of the Laurent series in $M \sim 0$ of the BFKL
kernel. Because of the singularities at $M= \pm 1$ this series has
radius of convergence one. Similarly the Laurent series for the 
anticollinear singularities around $M=1$ also has radius of convergence one.
Thus we expect the approximate calculations to do well over the central region
$0\leq M \leq 1$, but to break down as $M\to -1$, $M\to 2$. In 
figure $3$ we show how well the approximation
actually performs at LO and NLO, where the exact result is known. 
As expected the agreement is excellent close to
$M=0$ and $M=1$, and even in the central region the difference
between the collinear approximation and the full result is at the
percent level. Hence we can conclude that at leading twist
the collinear kernel is a very good approximation
to the full LO and NLO ones. For this reason we also expect our result
for $\chi_2$ to be a good approximation, within a few per cent, 
for calculations performed at leading twist.
A reasonable variation of the unknown contribution to 
the NLLx scheme change in our
calculation changes the kernel $\chi_2$ by $ \sim 5\%$,
hence well within the accuracy we expect for our approximation.

It is well known that beyond NLO BFKL evolution presents various unsolved
problems.  A direct computation  shows that the  universality of the
pomeron exchange is broken at NNLO \cite{delduca}.
 Furthermore a new class of contributions
involving the $t$-channel exchange of four gluons enters at
NNLO (see \cite{bartels} and references therein). These are
twist-four contributions which can mix with the twist-four part of
the two-gluon operators. The form of the full BFKL equation at
NNLO is thus different from that at LO and NLO, in
contrast to the GLAP equation which has the same form to all
orders in perturbation theory. Nevertheless collinear factorisation and
running coupling duality guarantee the existence of a universal
and factorised leading twist kernel for small-$x$ evolution \cite{2006},
valid in the approximation where all higher twist contributions are 
suppressed.
\begin{figure}\label{plot3}
\begin{center}
\includegraphics[width=10cm,height=7cm]{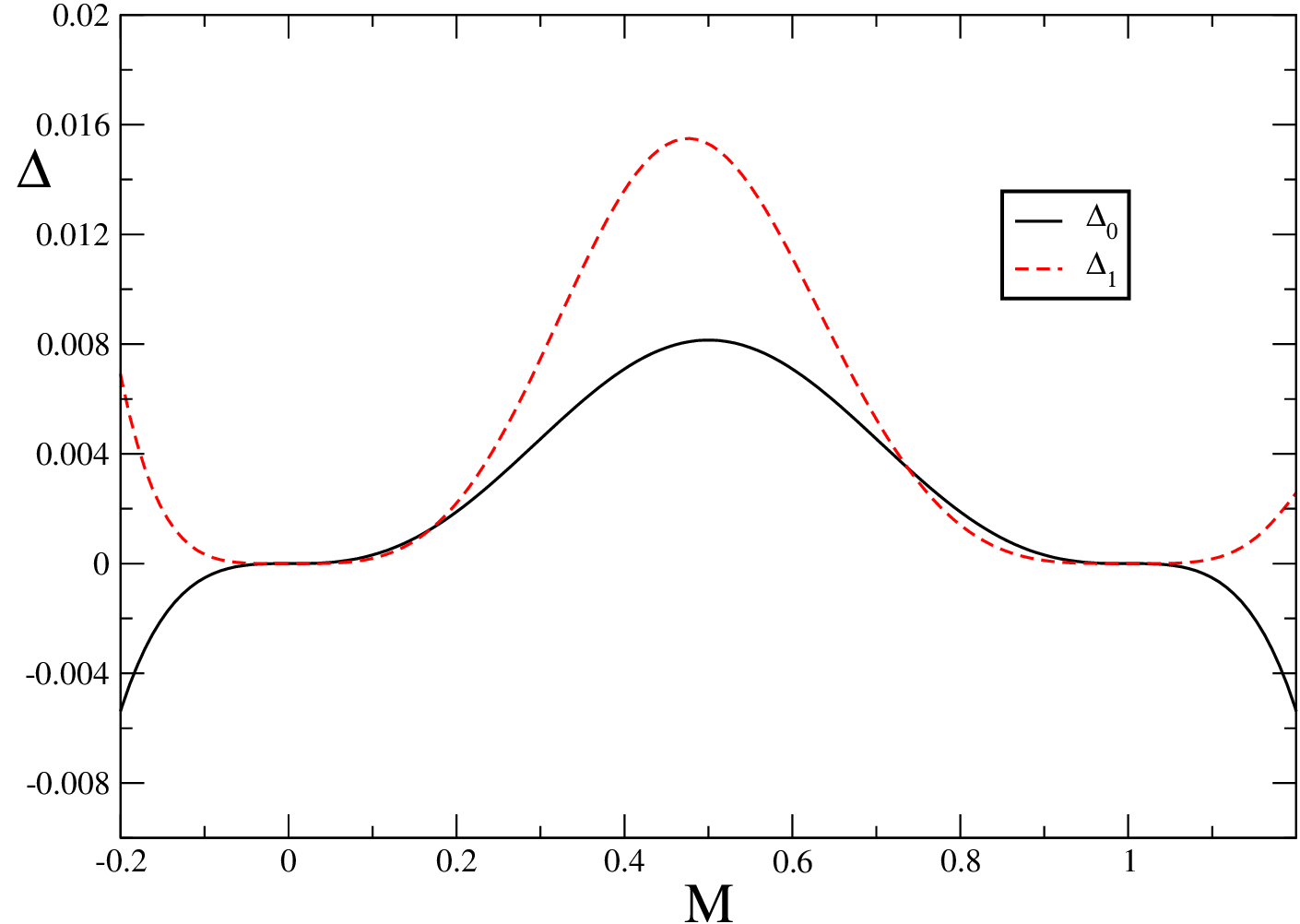}
\caption{This plot shows  the relative differences
$\Delta =$ \mbox{(exact-approximate)/exact} for the leading order kernel
($\Delta_0$) and the next-to-leading kernel ($\Delta_1$).}
\end{center}
\vskip-0.5truecm
\end{figure}

\section{Conclusions}
We have discussed the collinear approximation of the BFKL kernel.
Results on duality relations and  factorisation schemes, with the
inclusion of the running coupling enable us to construct
an approximation of the BFKL kernel at NNLO, which contains all
the singular contributions at $M=0$ and $M=1$. The collinear
approximation of $\chi_0$ and $\chi_1$ are in excellent agreement
with the full results and so our result for $\chi_2$ is also 
likely to be close to true result for the NNLO kernel in the region 
relevant at leading twist.

\begin{footnotesize}
\bibliographystyle{blois07}
{\raggedright
\bibliography{marzaniblois07}

\providecommand{\etal}{et al.\xspace}
\providecommand{\href}[2]{#2}
\providecommand{\coll}{Coll.}
\catcode`\@=11
\def\@bibitem#1{%
\ifmc@bstsupport
  \mc@iftail{#1}%
    {;\newline\ignorespaces}%
    {\ifmc@first\else.\fi\orig@bibitem{#1}}
  \mc@firstfalse
\else
  \mc@iftail{#1}%
    {\ignorespaces}%
    {\orig@bibitem{#1}}%
\fi}%
\catcode`\@=12
\begin{mcbibliography}{10}

\bibitem{chi1}
V.~S. Fadin and L.~N. Lipatov,
\newblock Phys. Lett.{} {\bf B429},~127~(1998).
\newblock \href{http://www.arXiv.org/abs/hep-ph/9802290}{{\tt
  hep-ph/9802290}}\relax
\relax
\bibitem{1997}
R.~D. Ball and S.~Forte,
\newblock Phys. Lett.{} {\bf B405},~317~(1997).
\newblock \href{http://www.arXiv.org/abs/hep-ph/9703417}{{\tt
  hep-ph/9703417}}\relax
\relax
\bibitem{2000}
G.~Altarelli, R.~D. Ball, and S.~Forte,
\newblock Nucl. Phys.{} {\bf B575},~313~(2000).
\newblock \href{http://www.arXiv.org/abs/hep-ph/9911273}{{\tt
  hep-ph/9911273}}\relax
\relax
\bibitem{vogt}
A.~Vogt, S.~Moch, and J.~A.~M. Vermaseren,
\newblock Nucl. Phys.{} {\bf B691},~129~(2004).
\newblock \href{http://www.arXiv.org/abs/hep-ph/0404111}{{\tt
  hep-ph/0404111}}\relax
\relax
\bibitem{salam}
G.~P. Salam,
\newblock Acta Phys. Polon.{} {\bf B30},~3679~(1999).
\newblock \href{http://www.arXiv.org/abs/hep-ph/9910492}{{\tt
  hep-ph/9910492}}\relax
\relax
\bibitem{1999}
R.~D. Ball and p.~Forte,
\newblock Phys. Lett.{} {\bf B465},~271~(1999).
\newblock \href{http://www.arXiv.org/abs/hep-ph/9906222}{{\tt
  hep-ph/9906222}}\relax
\relax
\bibitem{2002}
G.~Altarelli, R.~D. Ball, and S.~Forte,
\newblock Nucl. Phys.{} {\bf B621},~359~(2002).
\newblock \href{http://www.arXiv.org/abs/hep-ph/0109178}{{\tt
  hep-ph/0109178}}\relax
\relax
\bibitem{2006}
R.~D. Ball and S.~Forte,
\newblock Nucl. Phys.{} {\bf B742},~158~(2006).
\newblock \href{http://www.arXiv.org/abs/hep-ph/0601049}{{\tt
  hep-ph/0601049}}\relax
\relax
\bibitem{ciaf95}
M.~Ciafaloni,
\newblock Phys. Lett.{} {\bf B356},~74~(1995).
\newblock \href{http://www.arXiv.org/abs/hep-ph/9507307}{{\tt
  hep-ph/9507307}}\relax
\relax
\bibitem{ciaf05}
M.~Ciafaloni and D.~Colferai,
\newblock JHEP{} {\bf 09},~069~(2005).
\newblock \href{http://www.arXiv.org/abs/hep-ph/0507106}{{\tt
  hep-ph/0507106}}\relax
\relax
\bibitem{catani}
S.~Catani and F.~Hautmann,
\newblock Nucl. Phys.{} {\bf B427},~475~(1994).
\newblock \href{http://www.arXiv.org/abs/hep-ph/9405388}{{\tt
  hep-ph/9405388}}\relax
\relax
\bibitem{chi2}
S.~Marzani, R.~D. Ball, P.~Falgari, and S.~Forte~(2007).
\newblock \href{http://www.arXiv.org/abs/arXiv:0704.2404 [hep-ph]}{{\tt
  arXiv:0704.2404 [hep-ph]}}\relax
\relax
\bibitem{ciaf99}
M.~Ciafaloni, D.~Colferai, and G.~P. Salam,
\newblock JHEP{} {\bf 10}~(1999)\relax
\relax
\bibitem{delduca}
V.~Del~Duca and E.~W.~N. Glover,
\newblock JHEP{} {\bf 10},~035~(2001).
\newblock \href{http://www.arXiv.org/abs/hep-ph/0109028}{{\tt
  hep-ph/0109028}}\relax
\relax
\bibitem{bartels}
J.~Bartels and C.~Bontus,
\newblock Phys. Rev.{} {\bf D61},~034009~(2000).
\newblock \href{http://www.arXiv.org/abs/hep-ph/9906308}{{\tt
  hep-ph/9906308}}\relax
\relax
\end{mcbibliography}
}
\end{footnotesize}
\end{document}